\documentclass[aps,preprint,prl,amsmath,amssymb,showpacs,superscriptaddress]{revtex4-2}

\usepackage{CJK}
\usepackage{graphicx}% Include figure files
\usepackage{dcolumn}% Align table columns on decimal point
\usepackage{bm}
\usepackage{graphicx}
\usepackage{overpic}
\usepackage{longtable}
\usepackage{epsfig}
\usepackage{epsf}
\usepackage{booktabs}
\usepackage{colortbl}
\usepackage{booktabs}
\usepackage{amssymb}
\usepackage{amsmath}
\usepackage{amsthm}
\usepackage{multirow}
\usepackage{verbatim}
\usepackage[T1]{fontenc}
\usepackage{amsfonts}
\usepackage{scalerel}
\usepackage{cases}
\usepackage[colorlinks=true,linkcolor=blue,citecolor=blue,pdfauthor={ },pdftitle={ },pdfsubject={ },pdfkeywords={ }]{hyperref}
\usepackage{pgfplots}
\usepackage{float}
\usepackage{lipsum}
\usepackage{gensymb}

\newcommand{\myfont}{\fontfamily{phv}\selectfont}
\newcommand{\figfont}{\myfont \textsf \large}

\begin{document}

\title{Temperature-dependent behaviors of single spin defects in solids determined with Hz-level precision}

\author{Shaoyi Xu}
\thanks{S. Xu and M. Liu contributed equally to this work, from experimental and theoretical aspects, respectively.}
\affiliation{CAS Key Laboratory of Microscale Magnetic Resonance and School of Physical Sciences, University of Science and Technology of China, Hefei 230026, China}
\affiliation{CAS Center for Excellence in Quantum Information and Quantum Physics, University of Science and Technology of China, Hefei 230026, China}

\author{Mingzhe Liu}
\thanks{S. Xu and M. Liu contributed equally to this work, from experimental and theoretical aspects, respectively.}
\affiliation{CAS Key Laboratory of Microscale Magnetic Resonance and School of Physical Sciences, University of Science and Technology of China, Hefei 230026, China}
\affiliation{CAS Center for Excellence in Quantum Information and Quantum Physics, University of Science and Technology of China, Hefei 230026, China}

\author{Tianyu Xie}
\email{xie1021@ustc.edu.cn}
\affiliation{CAS Key Laboratory of Microscale Magnetic Resonance and School of Physical Sciences, University of Science and Technology of China, Hefei 230026, China}
\affiliation{CAS Center for Excellence in Quantum Information and Quantum Physics, University of Science and Technology of China, Hefei 230026, China}

\author{Zhiyuan Zhao}
\affiliation{CAS Key Laboratory of Microscale Magnetic Resonance and School of Physical Sciences, University of Science and Technology of China, Hefei 230026, China}
\affiliation{CAS Center for Excellence in Quantum Information and Quantum Physics, University of Science and Technology of China, Hefei 230026, China}

\author{Qian Shi}
\affiliation{CAS Key Laboratory of Microscale Magnetic Resonance and School of Physical Sciences, University of Science and Technology of China, Hefei 230026, China}
\affiliation{CAS Center for Excellence in Quantum Information and Quantum Physics, University of Science and Technology of China, Hefei 230026, China}

\author{Pei Yu}
\affiliation{CAS Key Laboratory of Microscale Magnetic Resonance and School of Physical Sciences, University of Science and Technology of China, Hefei 230026, China}
\affiliation{CAS Center for Excellence in Quantum Information and Quantum Physics, University of Science and Technology of China, Hefei 230026, China}

\author{Chang-Kui Duan}
\email{ckduan@ustc.edu.cn}
\affiliation{CAS Key Laboratory of Microscale Magnetic Resonance and School of Physical Sciences, University of Science and Technology of China, Hefei 230026, China}
\affiliation{CAS Center for Excellence in Quantum Information and Quantum Physics, University of Science and Technology of China, Hefei 230026, China}
\affiliation{Hefei National Laboratory, University of Science and Technology of China, Hefei 230088, China}

\author{Fazhan Shi}
\affiliation{CAS Key Laboratory of Microscale Magnetic Resonance and School of Physical Sciences, University of Science and Technology of China, Hefei 230026, China}
\affiliation{CAS Center for Excellence in Quantum Information and Quantum Physics, University of Science and Technology of China, Hefei 230026, China}
\affiliation{Hefei National Laboratory, University of Science and Technology of China, Hefei 230088, China}
\affiliation{School of Biomedical Engineering and Suzhou Institute for Advanced Research, University of Science and Technology of China, Suzhou 215123, China}

\author{Jiangfeng Du}
\email{djf@ustc.edu.cn}
\affiliation{CAS Key Laboratory of Microscale Magnetic Resonance and School of Physical Sciences, University of Science and Technology of China, Hefei 230026, China}
\affiliation{CAS Center for Excellence in Quantum Information and Quantum Physics, University of Science and Technology of China, Hefei 230026, China}
\affiliation{Hefei National Laboratory, University of Science and Technology of China, Hefei 230088, China}

\begin{abstract}
Revealing the properties of single spin defects in solids is essential for quantum applications based on solid-state systems. However, it is intractable to investigate the temperature-dependent properties of single defects, due to the low precision for single-defect measurements in contrast to defect ensembles. Here we report that the temperature dependence of the Hamiltonian parameters for single negatively charged nitrogen-vacancy (NV$^{-}$) centers in diamond is precisely measured, and the results find a reasonable agreement with first-principles calculations. Particularly, the hyperfine interactions with randomly distributed $^{13}$C nuclear spins are clearly observed to vary with temperature, and the relevant coefficients are measured with Hz-level precision. The temperature-dependent behaviors are attributed to both thermal expansion and lattice vibrations by first-principles calculations. Our results pave the way for taking nuclear spins as more stable thermometers at nanoscale. The methods developed here for high-precision measurements and first-principles calculations can be further extended to other solid-state spin defects.
\end{abstract}

\maketitle

\section{Introduction}
Accurate knowledge of the properties of spin defects in solids \cite{wolfowicz2021quantum} is the basis for finding their applications in quantum sensing \cite{degen2017quantum}, and quantum computation and networks \cite{awschalom2018quantum}. Measuring the susceptibilities of the target defect to external perturbations such as magnetic field, electric fields, strains, and temperature, enables the detection of these quantities and the analysis of the decoherence resulting from their fluctuations. As one of the most prominent systems, the nitrogen-vacancy (NV) center in diamond, with its various properties carefully investigated \cite{doherty2013nitrogen}, has acquired several remarkable achievements, including single-molecule magnetic resonance \cite{shi2015single, lovchinsky2016nuclear, shi2018single}, nanoscale magnetic \cite{thiel2019probing, ku2020imaging, song2021direct} and temperature \cite{kucsko2013nanometre, neumann2013high} imaging, and multi-node quantum networks \cite{pompili2021realization, hermans2022qubit}.

With regard to the temperature dependence of the NV properties, in early works, the zero-field splitting (ZFS) is found to be temperature-dependent \cite{acosta2010temperature, chen2011temperature, toyli2012measurement, doherty2014temperature}, which enables the NV center to work as a nanoscale thermometer \cite{kucsko2013nanometre, neumann2013high, toyli2013fluorescence}. Recently, the temperature dependence of the hyperfine interactions with the surrounding $^{14}$N and $^{13}$C nuclear spins is also explored based on NV ensembles \cite{barson2019temperature, soshenko2020temperature, jarmola2020robust, wang2022characterizing}, which can provide more information on the temperature dependence of the spin-density distribution of the NV$^{-}$ ground state. However, it is almost impossible to observe the temperature-dependent behaviors of the $^{13}$C hyperfine interactions for these NV-ensemble-based works, since the $^{13}$C atoms are randomly distributed in the proximity of the NV center.

In this work, we utilize single NV centers to investigate the temperature dependence of the parameters involved in the ground-state Hamiltonian of the NV$^{-}$ center. By performing Ramsey interferometry, the temperature dependence of the nearby $^{13}$C spins with the coupling strengths 13.7 MHz, 12.8 MHz, $-$8.9 MHz and $-$6.5 MHz is clearly observed, and the temperature coefficients are measured with Hz-level precision. Furthermore, first-principles calculations are performed based on density functional theory (DFT) \cite{gali2008ab, ranjbar2011many, maze2011properties, doherty2012theory, gali2019ab}, and the calculation results explain the experimental values fairly well. The temperature dependence of the hyperfine interactions is identified as the effects of both thermal expansion and lattice vibrations, which is different from the conclusion of the work \cite{tang2022first}. Our methods combining high-precision measurements and first-principles calculations are generally applicable for other defects in solids, such as phosphorus dopants in silicon \cite{mkadzik2022precision}, silicon vacancies in silicon carbide \cite{widmann2015coherent}, cerium ions in yttrium aluminium garnet \cite{siyushev2014coherent}, and ytterbium ions in yttrium orthovanadate \cite{ruskuc2022nuclear}.

\section{System and methods}
\begin{figure}[htbp]
\centering
\begin{overpic}[width=0.8\columnwidth]{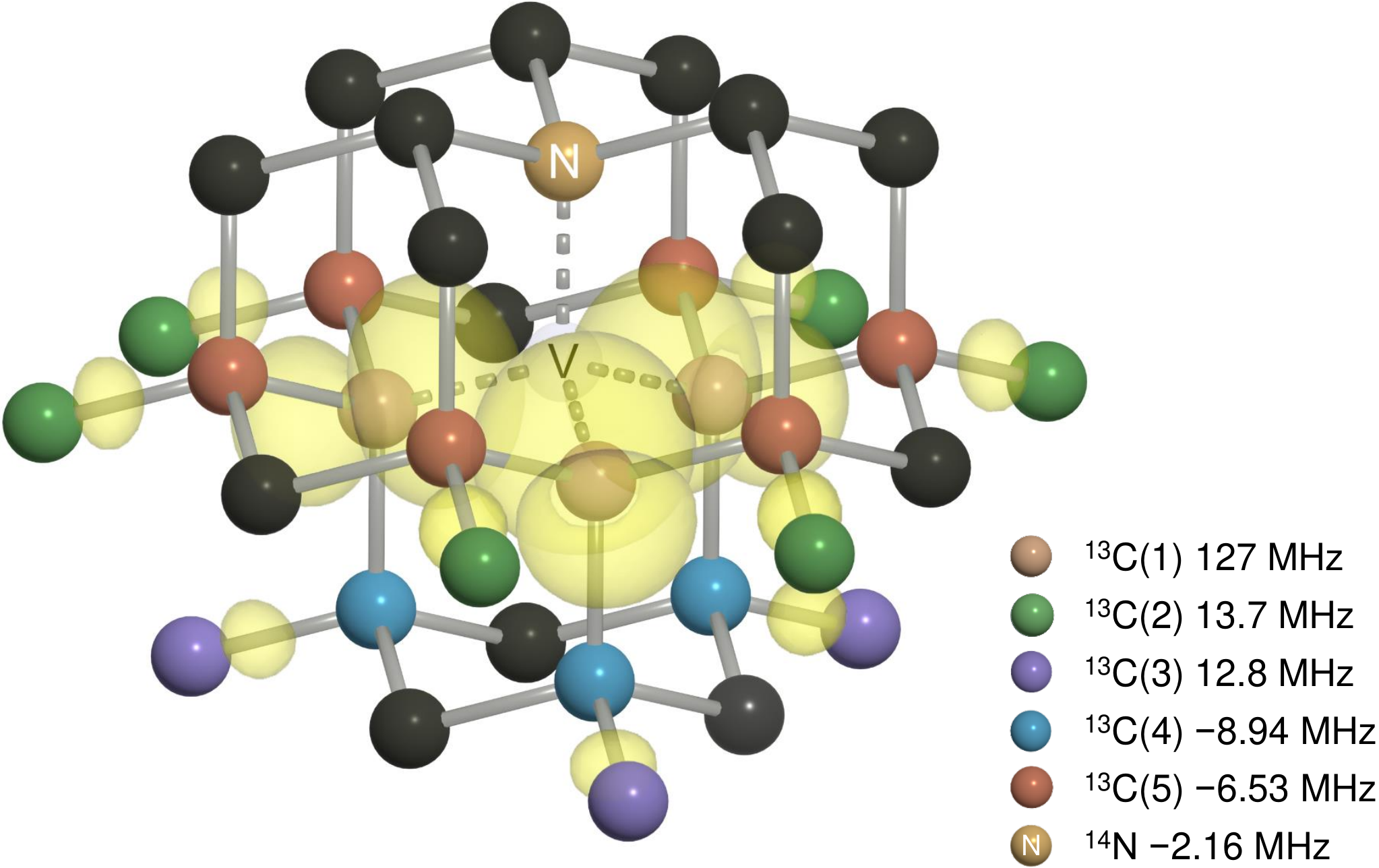}
\end{overpic}
\caption{Atomic structure of the NV center in diamond lattice and the calculated spin-density distribution. The dark yellow sphere denotes the nitrogen atom, while the white one represents the vacancy. The other spheres except the dark ones denote the carbon atoms studied in this work with the coupling strengths displayed in the lower right part. The spin density of the NV center spreads across multiple lattice sites, and interacts with the nearby $^{14}$N and $^{13}$C nuclear spins through magnetic dipolar moments. The distribution of the spin density is calculated based on DFT.
}\label{system}
\end{figure}

The NV center in diamond lattice consists of a substitutional $^{14}$N atom and an adjacent vacancy, as shown in Fig.~\ref{system}(a). The electronic state studied here is the ground state of the NV$^{-}$ spin triplet \cite{doherty2013nitrogen}. Two temperature-dependent phenomena in solids, i.e., thermal expansion and lattice vibrations, both have significant perturbations on the distribution of the ground-state waveform that determines the coupling parameters involved in the Hamiltonian concerning the NV electron spin and the nuclear spins. Thus, the nuclear spins, especially the $^{13}$C spins on multiple lattice sites (Fig.~\ref{system}(a)), can serve as atomic-scale sensors to probe the electron waveform and its variation with external perturbations, e.g., the temperature in this work. By constructing an optically detected magnetic resonance (ODMR) setup with temperature control \cite{xie2021beating} (see Supplemental Material \cite{sm}\nocite{childress2006coherent, jacques2009dynamic, steiner2010universal, blochl2000first, yazyev2005core, blochl1994projector, kresse1993ab, kresse1994ab, gyromagnetic, sato2002thermal}), the temperature dependence of the coupling parameters for single NV$^{-}$ centers can be investigated.

The bulk diamonds used here are all ultrapure with $^{13}$C natural abundance (see Supplemental Material \cite{sm}). Considering the hyperfine interactions with the $^{14}$N nuclear spin \cite{xie2021identity} and various $^{13}$C nuclear spins \cite{gali2009identification, smeltzer201113c}, the ground-state Hamiltonian under a bias field $\mathbf{B}$ with taking the NV axis as the $z$ direction can be formulated as
\begin{gather}
H_{g} = H_{e} + H_{\text{N}} + H_{\text{C}} \label{Hamiltonian_ground},\\
H_{e} = D(T) S_z^2 + \gamma_e \mathbf{B} \cdot \mathbf{S} \label{Hamiltonian_electron},\\
H_{\text{N}} = P(T) (I_z^{\text{N}})^2 - \gamma_n^{\text{N}} \mathbf{B} \cdot \mathbf{I}^{\text{N}} + \mathbf{S} \cdot \mathbf{A}^{\text{N}}(T) \cdot \mathbf{I}^{\text{N}} \label{Hamiltonian_N},\\
H_{\text{C}} = - \gamma_n^{\text{C}} \sum_i \mathbf{B} \cdot \mathbf{I}_i^{\text{C}} + \mathbf{S} \cdot \sum_i \mathbf{A}_i^{\text{C}}(T) \cdot \mathbf{I}_i^{\text{C}} \label{Hamiltonian_C},
\end{gather}
where the temperature-dependent parameters include the ZFS of the NV spin $D(T)$, the quadrupole coupling $P(T)$ of the $^{14}$N spin, and the hyperfine interaction $\mathbf{A}^{\text{N}}(T)$ of the $^{14}$N spin and $\mathbf{A}_i^{\text{C}}(T)$ of the $^{13}$C$(i)$ spin. $\mathbf{S}$, $\mathbf{I}^{\text{N}}$, and $\mathbf{I}_i^{\text{C}}$ are the operators of the NV spin, the $^{14}$N spin, and the $^{13}$C$(i)$ spin, respectively. $\gamma_e$, $\gamma_n^{\text{N}}$, and $\gamma_n^{\text{C}}$ are the gyromagnetic ratios of three kinds of spins. The coupling tensor $\mathbf{A}^{\text{N}}(T)$ only has two independent parameters due to the $C_{3v}$ symmetry, while the $\mathbf{A}_i^{\text{C}}(T)$ has six.

In this work, the temperature-dependent parameters described above can all be precisely determined by measuring the transition frequencies of the electron spin and the nuclear spins. First, the ZFS of the electron spin $D(T)$ in Eq.~(\ref{Hamiltonian_electron}) can be easily obtained with kHz-level precision by performing pulsed ODMR spectra. Second, by using the method in \cite{xie2021identity}, the quadrupole coupling $P(T)$ and the hyperfine interaction $\mathbf{A}^{\text{N}}(T)$ in Eq.~(\ref{Hamiltonian_N}) can both be solved out by measuring six nuclear transition frequencies with Hz-level precision under a field of $\sim$ 510 G. As for the $^{13}$C hyperfine interactions, although the tensor $\mathbf{A}_i^{\text{C}}(T)$ in Eq.~(\ref{Hamiltonian_C}) cannot be fully solved out due to its complexity, the temperature dependence can still be obtained by averaging two nuclear transition frequencies under a small bias field of 10$-$30 G with
\begin{equation}
\label{13C_hyperfine}
A = \frac{1}{2}(\omega_{+1}+\omega_{-1}) = \sqrt{A_{zx}^2+A_{zy}^2+A_{zz}^2} + R,
\end{equation}
where $\omega_{+1}$ and $\omega_{-1}$ are the transition frequencies of the $^{13}$C spin in both $m_{\scaleto{S}{3.4pt}} = +1$ and $m_{\scaleto{S}{3.4pt}} = -1$ subspaces of the NV spin. The remainder term $R$ (see Supplemental Material \cite{sm}) is constant with the temperature if the bias field is stable enough, and thus measuring the mean $A$ under different temperatures gives the temperature coefficient of the coupling term $\sqrt{A_{zx}^2+A_{zy}^2+A_{zz}^2}$.

\section{Experiments}
In the following, the temperature dependence of the relevant parameters is measured experimentally based on the discussions above. At first, by performing pulsed ODMR spectra under different temperatures, the variation of the ZFS  with the temperature is obtained to be $-$71.9(0.3) kHz/K for single NV centers at room temperature (see Supplemental Material \cite{sm}). The deviation from the ensemble-NV result $-$74.2(0.7) kHz/K \cite{acosta2010temperature} may be originated from the vast strain difference between the diamond samples used in two works or the systematic error for temperature measurements. Then, by applying the method \cite{xie2021identity} for measuring the quadrupole coupling $P(T)$ and the hyperfine interaction $\mathbf{A}^{\text{N}}(T)$ of the $^{14}$N nuclear spin, the temperature coefficients are given by 35.0(0.3) Hz/K and 194.9(1.0) Hz/K (see Supplemental Material \cite{sm}) in a good agreement with the previous ensemble-NV results \cite{soshenko2020temperature, jarmola2020robust, wang2022characterizing}.

\begin{figure*}[htbp]
\centering
\begin{overpic}[width=1.0\columnwidth]{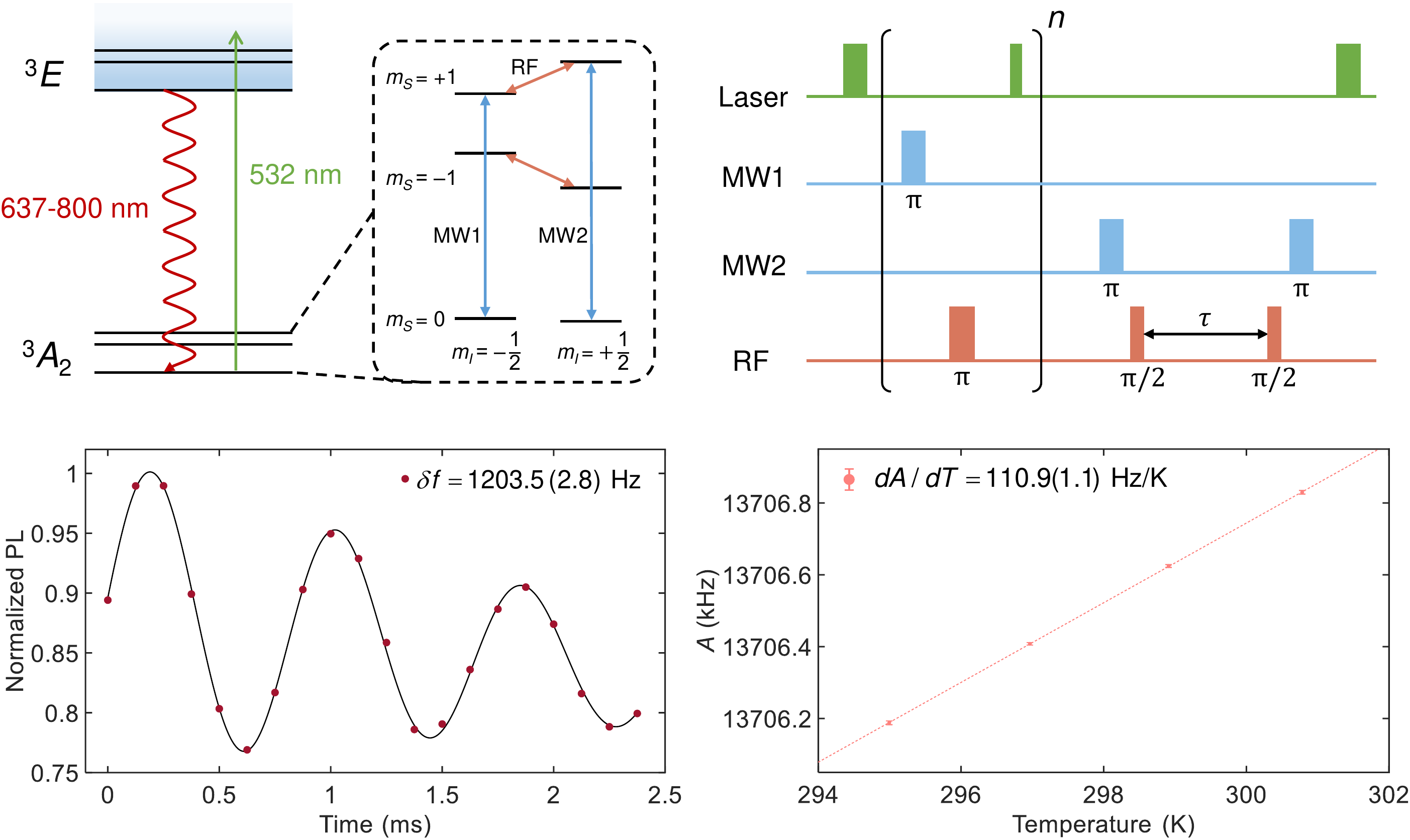}
	\put (-0.2, 58.5) {\figfont{(a)}}
	\put (48.5, 58.5) {\figfont{(b)}}
	\put (-0.2, 28.5) {\figfont{(c)}}
	\put (48.5, 28.5) {\figfont{(d)}}
\end{overpic}
\caption{Measurement for the temperature dependence of the hyperfine interaction of a $^{13}$C(2) nuclear spin. (a) Level diagram for the NV center strongly coupled to a $^{13}$C(2) nuclear spin. The 532-nm laser pulse is used to initialize the NV electron spin into the state $|m_{\scaleto{S}{3.4pt}} = 0\rangle$, and readout the spin state by collecting fluorescence photons with the 637$-$800 nm phonon sideband. The orange arrows indicate the two nuclear transitions with the frequencies $\omega_{+1}$ and $\omega_{-1}$ in Eq.~(\ref{13C_hyperfine}) to be measured, which are driven by radiofrequency (RF) pulses. The transitions for the NV electron spin, as indicated by the blue arrows, are driven by microwave (MW) pulses. The RF, MW1, and MW2 pulses are used in the pulse sequence in (b). (b) The pulse sequence of laser, MW, and RF for the Ramsey interference between the state $|m_{\scaleto{S}{3.4pt}} = +1, m_{\scaleto{I}{3.4pt}} = +\frac{1}{2}\rangle$ and the state $|m_{\scaleto{S}{3.4pt}} = +1, m_{\scaleto{I}{3.4pt}} = -\frac{1}{2}\rangle$. The sequence enclosed by the brackets is repeated $n$ times to polarize the $^{13}$C(2) spin into the state $|m_{\scaleto{I}{3.4pt}} = +\frac{1}{2}\rangle$. (c) Resultant interference pattern after applying the sequence in (b). The black line is plotted by fitting the data with the function $\{a\sin[2\pi(\delta f)t + \phi_0] + b\}\exp[-(t/T_2^*)^p] + c$, giving the detuning $\delta f = $ 1203.5(2.8) Hz. (d) The mean $A$ of the two nuclear transition frequencies $\omega_{+1}$ and $\omega_{-1}$ measured under different temperatures. The temperature coefficient of the $^{13}$C(2) nuclear spin at room temperature is given by 110.9(1.1) Hz/K with a linear fit.
}\label{13C}
\end{figure*}

The main challenge in the experiments is to measure the temperature dependence of the hyperfine interactions $\mathbf{A}^{\text{C}}(T)$ for the $^{13}$C nuclear spins in the proximity of single NV centers. Before the measurement, a small bias field of 10$-$30 G is applied and aligned to the NV axis by adopting the method of three-level quantum beat \cite{shim2013characterization}. Figure \ref{13C} shows the measurement process by taking a $^{13}$C(2) spin (Fig.~\ref{system}(a)) as an example. Based on the level structure of the NV$-^{13}$C(2) coupled system shown in Fig.~\ref{13C}(a), the Ramsey sequence together with that for polarizing the $^{13}$C(2) spin, as displayed in Fig.~\ref{13C}(b), is applied for measuring the $^{13}$C(2) transition frequency $\omega_{+1}$ in the $m_{\scaleto{S}{3.4pt}} = +1$ subspace of the NV spin. The resulting interference pattern is plotted in Fig.~\ref{13C}(c) with the fitting curve. The value of 13684603.5(2.8) Hz for $\omega_{+1}$ is obtained by adding the detuning $\delta f$ to the RF frequency used in the Ramsey sequence. By repeating the process above to acquire the transition frequencies $\omega_{+1}$ and $\omega_{-1}$ under different temperatures, the temperature coefficient for the $^{13}$C(2) spin is given by 110.9(1.1) Hz/K, as shown in Fig.~\ref{13C}(d).

\begin{figure}[b]
\centering
\begin{overpic}[width=0.8\columnwidth]{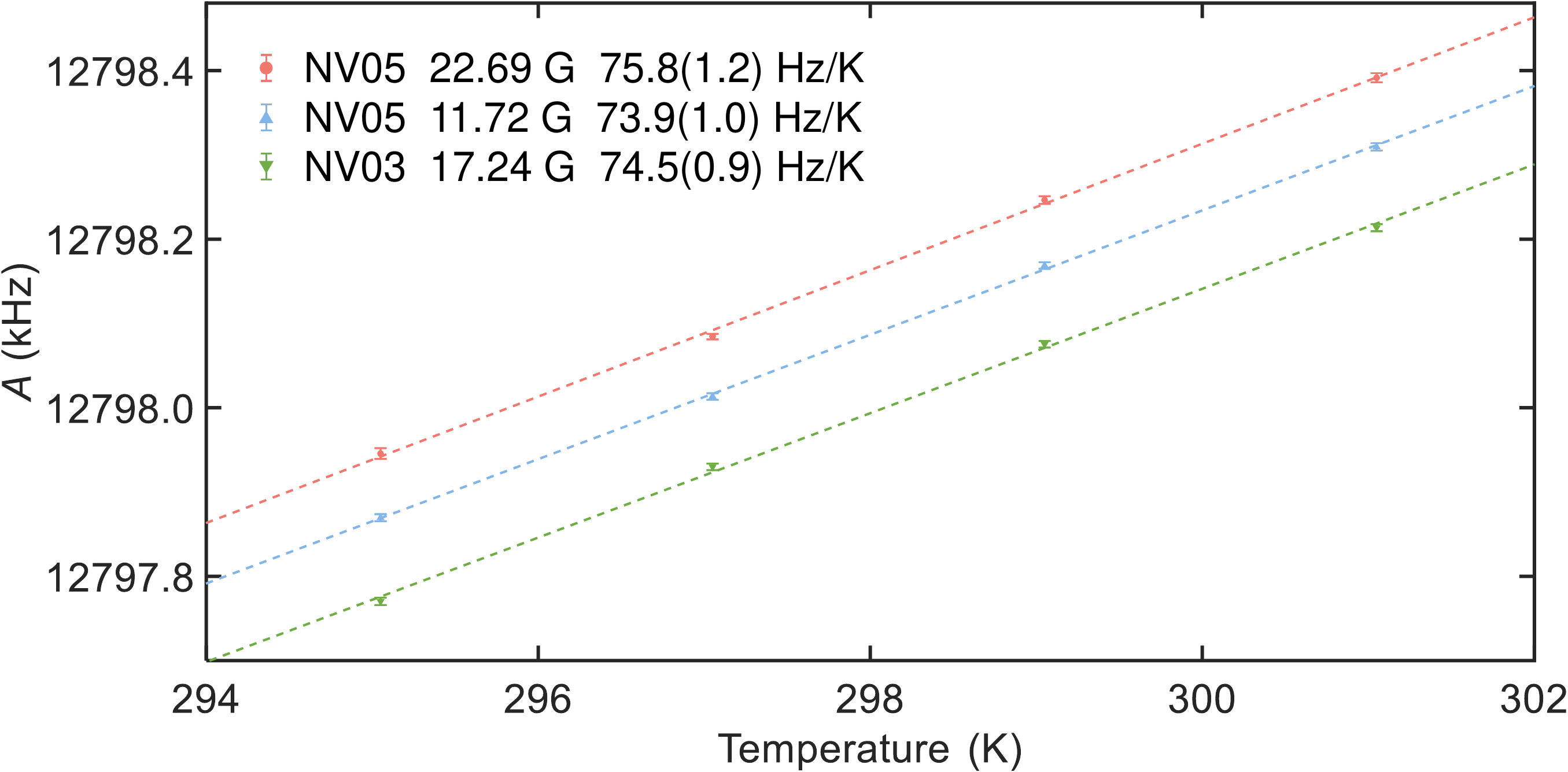}
\end{overpic}
\caption{The experiments performed under different bias fields. The temperature coefficient of the hyperfine interaction for $^{13}$C(3) nuclear spins is measured under three bias fields for two NV centers.
}\label{validity}
\end{figure}

There are some other effects induced by varying the temperature inside the box, e.g., the drift of the bias field, and these effects may disturb the measurement results above. Therefore, in order to ensure that the measured temperature-dependent behaviors are indeed originated from the temperature dependence of the $^{13}$C hyperfine interactions, the same experiments in Fig.~\ref{13C} are implemented under three bias fields for two NV centers that are coupled to $^{13}$C(3) nuclear spins. The results shown in Fig.~\ref{validity} are identical within the error bars, which verifies the validity and robustness of the measurement method adopted in this work. The final temperature coefficients for four kinds of $^{13}$C spins and the $^{14}$N spin are given by averaging the results of fifteen NV centers in four diamond samples (see Supplemental Material \cite{sm}), and summarized in Fig.~\ref{cal_figs}(d).

\section{First-principles calculations}
Here we perform first-principles calculations to find reasonable explanations for the experimental results above. The coupling tensor $\mathbf{A}$ for a nuclear spin can be calculated by averaging the magnetic dipolar interaction \cite{jackson1999classical} over the spin-density distribution of the NV$^{-}$ ground state, which includes the isotropic Fermi contact term and the anisotropic dipolar term. In our first-principles calculations, $4\times 4\times 4$ supercells are adopted to represent the NV$^{-}$ centers. The vibration modes at the $\Gamma$ point and the electron-spin densities under different geometric structures are calculated with the Perdew-Burke-Ernzerhof (PBE) density functional \cite{perdew1996generalized} and an energy cutoff of 400 eV. Here we are focusing on the calculation of the coupling term $A=\sqrt{A_{zx}^2+A_{zy}^2+A_{zz}^2}$ for directly comparing with the experimental results. On its temperature dependence, the contributions from thermal expansion and lattice vibrations, denoted as the static part $\delta A_{\rm stc}(T)$ and the dynamic part $\delta A_{\rm dyn}(T)$, are separately treated as small corrections,
\begin{equation}
A(T)=A(0)+\delta A_{\rm stc}(T)+\delta A_{\rm dyn}(T).
\label{A(T)_expression}
\end{equation}

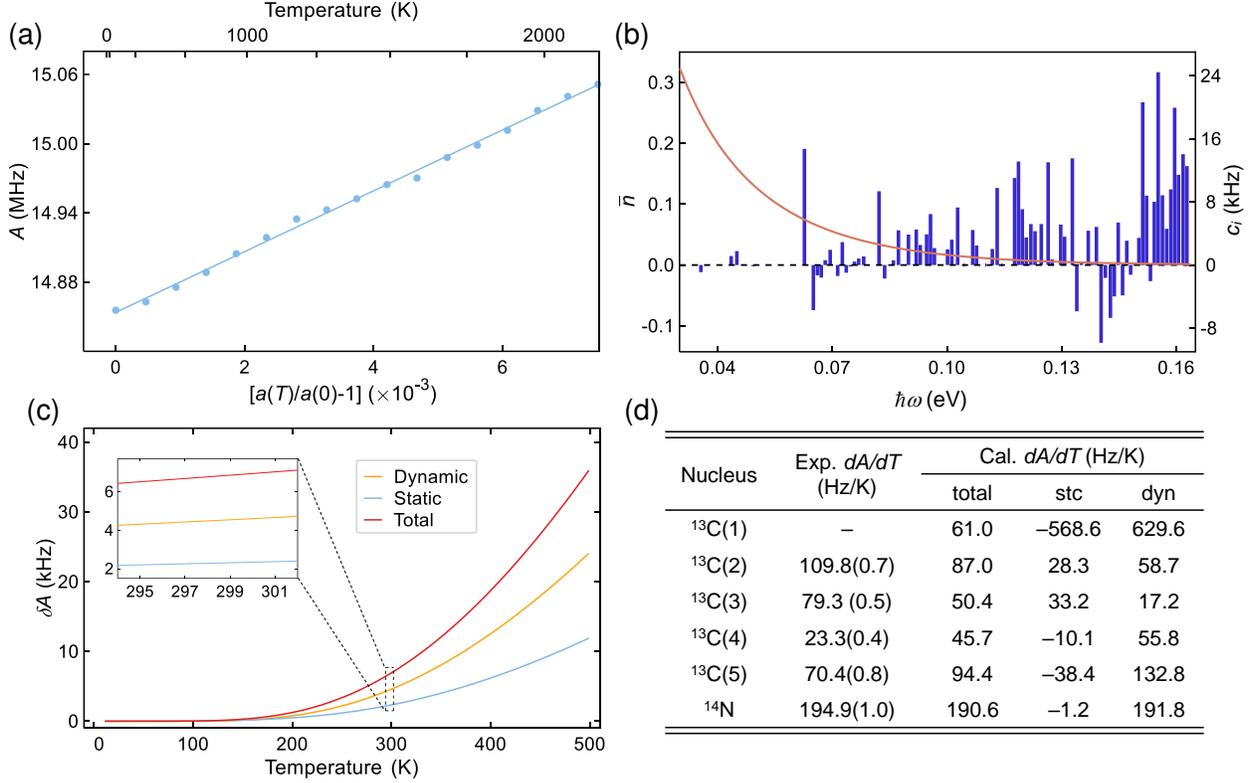
\begin{figure*}[t]
\centering
\begin{overpic}[width=1.0\columnwidth]{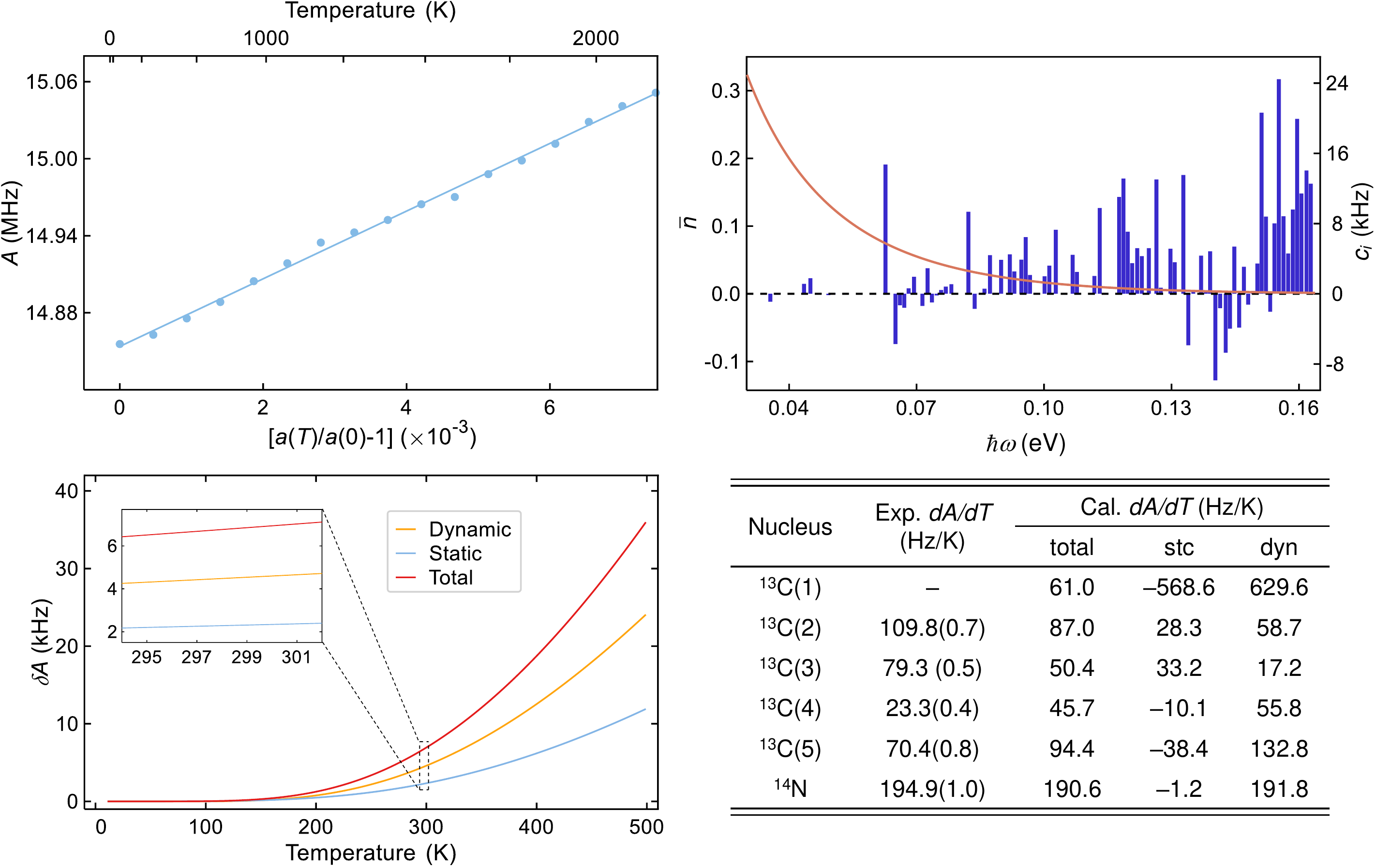}
	\put (0., 59.6) {\figfont{(a)}}
	\put (49.0, 59.6) {\figfont{(b)}}
	\put (1.5, 29.2) {\figfont{(c)}}
	\put (49.8, 29.2) {\figfont{(d)}}
\end{overpic}
\caption{Calculation of the temperature-dependent coupling $A(T)$. (a) The coupling strength $A=\sqrt{A_{zx}^2+A_{zy}^2+A_{zz}^2}$ for the $^{13}$C(2) spin under the thermal expansion $[a(T)/a(0)-1]$ with the corresponding temperature given by the upper abscissa axis, where the tiny tick near 0 K is 250 K. It clearly shows that the static contribution $\delta A_{\rm stc}(T)$ is proportional to the change of the lattice constant. (b) The average phonon number $\bar n(T)=[\exp(\hbar\omega/k_{\rm B}T)-1]^{-1}$ at $T=300$ K, and the dynamical contribution per phonon $c_i$ for the $^{13}$C(2) spin as a function of the phonon energy $\hbar\omega$. It is noted that the $c_i$'s from nearly degenerate modes (within $1~$meV) are merged for display. (c) The total thermal correction $\delta A(T)$ and its composition for the $^{13}$C(2) spin. Both static (i.e., thermal expansion) and dynamic (i.e., lattice vibrations) contributions are significant. (d) Comparison between the calculations and the experimental results at room temperature on the temperature dependence of the parameters regarding the nearby nuclear spins. The last two columns represent the contributions from thermal expansion (stc) and lattice vibrations (dyn). All errors in parentheses stand for one standard deviation.
}\label{cal_figs}
\end{figure*}

The static part $\delta A_{\rm stc}(T)$ is obtained by considering the impact of the temperature-related lattice expansion, with all atoms in the supercell residing at their static equilibrium positions. Our calculations show that $\delta A_{\rm stc}(T)$ is proportional to the expansion of the lattice constant $[a(T)/a(0)-1]$ \cite{jacobson2019thermal} (see Supplemental Material \cite{sm} for details),
\begin{equation}
\delta A_{\rm stc}= c_{\rm stc} \left [\frac{a(T)}{a(0)}-1\right ].
\label{A_stc_temp}
\end{equation}
The calculation results of $^{13}$C(2) spin are plotted in Fig.~\ref{cal_figs}(a). To reduce the impact of numerical errors in the first-principles calculations, a much larger range of $[a(T)/a(0)-1]$ than 0$-$300 K is adopted to obtain the coefficient $c_{\rm stc}$.

To obtain $\delta A_{\rm dyn}(T)$, $A_{\rm stc}(X_i)$ is calculated as a function of the canonical coordinate $X_i$ for the vibration mode $i=1$$-$1530 of the supercell. The function $A_{\rm stc}(X_i)$ can be well fitted by a rank-2 polynomial as follows (see Supplemental Material \cite{sm} for details):
\begin{equation}
A_{\rm stc}(X_i) - A_{\rm stc}(0) = b_i X_i + c_i \frac{\omega_i} {\hbar}X_i^{2},
\label{A_stc_coor}
\end{equation}
where $b_i$ and $c_i$ are the fitting parameters, $\omega_i$ is the phonon frequency, and $\hbar$ is the reduced Planck's constant. Evaluating and summing the thermodynamic expectation value of $A_{\rm stc}(X_i)- A_{\rm stc}(0)$ at the temperature $T$ over all the vibration modes, gives the expressions of $\delta A_{\rm dyn}(T)$ and $A(0)$ in Eq.~\eqref{A(T)_expression}
\begin{eqnarray}
\label{Adyn}
&&\delta A_{\rm dyn}(T)= \sum_i c_i {\bar{n}_i}(T),\\
&&A(0) = A_{\rm stc}(0) + \sum_i \frac{c_i}{2},
\label{Astatic_cor}
\end{eqnarray}
where  $\bar{n}_i(T) = [\exp(\hbar\omega_i/k_{\rm B}T)-1]^{-1}$ is the average phonon number of the vibration mode $i$ with $k_{\rm B}$ the Boltzmann constant, and the fitting parameter $c_i$ (defined in Eq.~\eqref{A_stc_coor}) gives the contribution per phonon to $\delta A_{\rm dyn}$. Considering the anharmonic effect, the first-order term in Eq.~\eqref{A_stc_coor} may contribute but is related to the static contribution included in Eq.~\eqref{A_stc_temp}. The cross second-order term $X_i X_j$ ($i\neq j$) involving the modes $i$ and $j$ are of higher order than those considered. The contributions of different vibration modes and the temperature-dependent part of $A(T)$ for the $^{13}$C(2) spin are displayed in Fig.~\ref{cal_figs}(b,c).

Figure \ref{cal_figs}(d) lists the temperature derivatives of $A(T)$ at 300 K for the $^{14}$N and $^{13}$C(1$-$5) spins. At 300 K, the dynamical part dominates for the ${}^{14}{\rm N}$ spin, while both the static and dynamical terms contribute significantly to various $^{13}$C spins (see Supplemental Material \cite{sm} for more details). Furthermore, the Fermi contact term always dominates over the anisotropic term for both the static part (Fig.~\ref{cal_figs}(a)) and the dynamical part (Fig.~\ref{cal_figs}(c)). Besides, the temperature-dependent terms are always much smaller than their corresponding $A(0)$ in magnitude ($\sim 1\%$ for the $^{14}$N spin and the scale of $0.01$$-$$0.1\%$ for the $^{13}$C spins at 300 K), which confirms the weak coupling between the lattice deformation and the hyperfine interaction.

\section{Conclusions}
In conclusion, the temperature coefficients of the temperature-dependent parameters contained in the ground-state Hamiltonian of single NV$^{-}$ centers are precisely measured at room temperature, and first-principles calculations explain the experimental results fairly well. Especially, the temperature coefficients of the quadrupole coupling of the $^{14}$N nuclear spin and the hyperfine interactions of the $^{14}$C and $^{13}$C spins are measured with Hz-level precision by performing Ramsey interferometry on the nuclear spins. Among these parameters, the hyperfine coupling of the $^{14}$N spin has the largest susceptibility to the temperature. Thus, it may work as a nanoscale thermometer like the ZFS \cite{kucsko2013nanometre, neumann2013high, toyli2013fluorescence}, considering that millisecond-scale coherence times can nearly remedy the gap in the temperature coefficient compared to the ZFS.

In the future, it is worthwhile to perform the measurements with varying strains or a wider range of temperatures for allowing a more detailed test of the calculation results. The calculations can be further improved by adopting a more accurate description of the NV$^{-}$ such as a larger supercell and a more accurate density functional, a higher-precision response of the variation of the spin-density function to tiny structure changes, and a more thorough description of the anharmonic effect. The methods for high-precision measurements and first-principles calculations in this work are universal, and can help deepen our understandings of NV centers as well as other solid-state defects \cite{mkadzik2022precision, widmann2015coherent, siyushev2014coherent, ruskuc2022nuclear}.

%\section*{Conflict of interest}
%The authors declare that they have no competing interests. 

\section*{Acknowledgements}
This work was supported by the National Key R\&D Program of China (Grant No.~2018YFA0306600, 2016YFA0502400), the National Natural Science Foundation of China (Grant No.~81788101, 91636217, T2125011, 12274396), Innovation Program for Quantum Science and Technology (Grant No.~2021ZD0302200, 2021ZD0303204), the CAS (Grant No.~XDC07000000, GJJSTD20200001, QYZDY-SSW-SLH004, Y201984), the Anhui Initiative in Quantum Information Technologies (Grant No.~AHY050000), the CAS Project for Young Scientists in Basic Research, the Fundamental Research Funds for the Central Universities, the China Postdoctoral Science Foundation (Grant No.~2021M703110, 2022T150631), and the Hefei Comprehensive National Science Center. The numerical calculations were performed on the supercomputing system at the Supercomputing Center of the University of Science and Technology of China.

%\section*{Author contributions}
%J.D. and F.S. supervised the project, and T.X. proposed the idea. T.X. designed the experiments, and S.X. implemented the experiments. C.D. and M.L. performed the first-principles calculations. S.X. and Z.Z. prepared the setup. P.Y. prepared the diamond samples. T.X., C.D., M.L., and S.X. wrote the manuscript. All authors analyzed the data, discussed the results and commented on the manuscript.

\bibliography{references}

\end{document}